\documentclass[aps, twocolumn, superscriptaddress, longbibliography]{revtex4-2}
\usepackage{times}
\usepackage{amsmath}
\usepackage{amssymb}
\usepackage{graphicx}
\usepackage[caption=false, position=top, singlelinecheck=off, justification=raggedright]{subfig}
\usepackage{color}
\usepackage[unicode]{hyperref}
\hypersetup{unicode=true, colorlinks=true, linkcolor=blue, citecolor=blue}
\usepackage{float}
\usepackage[normalem]{ulem}
\usepackage{multirow}
\usepackage{hhline}
\usepackage{mathrsfs,braket}
\usepackage{amssymb, amsbsy, amsmath, latexsym, dsfont, array, layout, graphicx, mathrsfs,ulem, color, bm}

\begin{document}
\title{Anomalous transport in quasiperiodic lattices: emergent exceptional points at band edges and log-periodic oscillations}
\author{Jinyuan Shang}
\affiliation{Beijing National Laboratory for Condensed Matter Physics, Institute of Physics, Chinese Academy of Sciences, Beijing 100190, China}
\affiliation{School of Physical Sciences, University of Chinese Academy of Sciences, Beijing 100049, China}
\author{Haiping Hu}
\email{hhu@iphy.ac.cn}
\affiliation{Beijing National Laboratory for Condensed Matter Physics, Institute of Physics, Chinese Academy of Sciences, Beijing 100190, China}
\affiliation{School of Physical Sciences, University of Chinese Academy of Sciences, Beijing 100049, China}

\begin{abstract}
Quasiperiodic systems host exotic transport regimes that are distinct from those found in periodic or disordered lattices. In this work, we study quantum transport in the Aubry-André-Harper lattice in a two-terminal setup coupled to zero-temperature reservoirs, where the conductance is evaluated via the nonequilibrium Green's function method. In the extended phase, we uncover a universal subdiffusive transport when the bath chemical potential aligns with the band edges. Specifically, the typical conductance displays a scaling of $\mathcal{G}_{\text{typ}}\sim L^{-2}$ with system size $L$. We attribute this behavior to the emergence of an exceptional point (Jordan normal form) in the transfer matrix in the thermodynamic limit. In the localized phase, the conductance shows exponential decay governed by the Lyapunov exponent. Intriguingly, in the critical phase, we identify pronounced log-periodic oscillations of the conductance as a function of system size, arising from the discrete scale invariance inherent to the singular-continuous spectrum. We further extend our analysis to the generalized Aubry-André-Harper model and provide numerical evidence suggesting that the exact mobility edge resides within a finite spectral gap. This results in a counter-intuitive exponential suppression of conductance precisely at the mobility edge. Our work highlights the distinct transport behaviors in quasiperiodic systems and elucidates how they are rigorously dictated by the underlying local spectral structure.
\end{abstract}
\maketitle
\section{Introduction}
Quasiperiodic lattices are deterministic aperiodic structures that interpolate between periodic crystals and disordered media. They can be engineered simply via incommensurate modulations of onsite energies, and have been realized, for example, with ultracold atoms in bichromatic optical lattices~\cite{Roati2008,Schreiber2015,Luschen2018,Wang2022MobilityEdge}. Over the past decades, quasiperiodic lattices have served as a rich platform for exploring physics beyond the periodic-disordered dichotomy, including localization transitions~\cite{Aubry1980,Harper1955,DasSarma1988,Wang2020,Biddle2010,Lahini2009,Ganeshan2015,Xu2020,ZhangYu2022,ZhouXC2023,ZhouXC2025,Boers2007,Biddle2009,DasSarma1990}, topological phases~\cite{Verbin2013,Verbin2015,Kraus2012a,Kraus2012b,Lang2012a,Lang2012b,Hu2015,Cai2013}, and the spectral theory of almost-Mathieu operators~\cite{Avila2009,Simon1982,Jitomirskaya1999,Avila2015,Avila2017,Jitomirskaya1994}. Among them, quantum transport in quasiperiodic lattices is a fundamental problem in its own right and is of particular interest. While perfect crystals typically support ballistic transport and disordered media suppress it via Anderson localization~\cite{Anderson1958,Abrahams1979,LeeRamakrishnan1985}, quasiperiodic systems can exhibit anomalous transport regimes, including subdiffusion~\cite{Hiramoto,Ketzmerick1997,Hiramoto,Purkayastha2017,Purkayastha2018,Varma2017,Sutradhar2019,ZhangYu2025} and even superdiffusion~\cite{Karrasch,Swingle,Purkayastha2018}, owing to their intricate spectral and eigenstate properties.

Unlike systems with random disorder, quasiperiodic lattices can possess multiple single-particle phases (e.g., extended, localized, and critical), each with distinct spectral and eigenstate characteristics, and may also support mobility edges \cite{DasSarma1990,Biddle2010,Ganeshan2015,Wang2020,Boers2007}. A standard way to probe transport is to attach two leads to the ends of the lattice. The leads act as zero-temperature reservoirs, and a small chemical-potential bias between them drives the system into a nonequilibrium steady state at long times~\cite{DharSen2006}. The steady-state conductance $\mathcal{G}(E,L)$ in this setup~\cite{Landauer1970,Economou1981} depends on the energy $E$ (set by the reservoir chemical potential) and the system size $L$. Transport is then characterized by the finite-size scaling of $\mathcal{G}(E,L)$ with $L$: ballistic transport corresponds to $\mathcal{G}\propto L^{0}$, and localization to $\mathcal{G}\propto e^{-\beta L}$. More generally, a power law $\mathcal{G}\propto L^{-\alpha}$ corresponds to superdiffusive, diffusive, and subdiffusive transport for $0<\alpha<1$, $\alpha=1$, and $\alpha>1$, respectively. While transport in quasiperiodic systems has been extensively studied, much of the existing work focuses on integrated observables, such as the steady current at high (or even infinite) temperature~\cite{Purkayastha2017,Purkayastha2018,Varma2017}, which inherently average over the entire spectrum. A microscopic understanding of how the spectral structure of quasiperiodic lattices dictates transport scaling at fixed energy remains elusive. In particular, the energy-resolved conductance $\mathcal{G}(E,L)$ provides a finer characterization of transport and allows one to ask how transport changes as the chemical potential $E$ is scanned across different spectral regimes and phases. 

In this work, we provide a comprehensive study of quantum transport in quasiperiodic lattices by analyzing the energy-resolved conductance $\mathcal{G}(E,L)$. We focus on the paradigmatic Aubry–André–Harper (AAH) model \cite{Aubry1980,Harper1955} and its generalized version \cite{Ganeshan2015}, which hosts a mobility edge. We systematically examine how $\mathcal{G}(E,L)$ behaves across different phases and spectral regimes. For the AAH model, we find that in the extended phase, transport becomes universally subdiffusive at the band edges on large length scales, with $\alpha=2$. We show that this behavior is tied to an emergent exceptional point in the transfer matrix. In the localized phase, the conductance decays exponentially with system size, with a decay rate set by the Lyapunov exponent. At the critical point, we demonstrate that the conductance deviates from a simple monotonic scaling with system size. Instead, it exhibits pronounced log-periodic oscillations, which act as a hallmark of the discrete scale symmetry inherent to the singular continuous spectrum~\cite{Sornette1998,Gluzman2002,Derrida1984,Nauenberg1975}. The period of these oscillations depends on the energy $E$ and is linked to the specific inflation factor encoded in the symbolic code. We then extend our analysis to the generalized AAH (GAAH) model. We again observe the same $\alpha=2$ scaling at band edges in the extended part of the spectrum. Intriguingly, when the chemical potential is tuned to the mobility edge, the conductance exhibits an unexpected exponential decay rather than algebraic scaling. This behavior is consistent with our numerical evidence of a finite, albeit small spectral gap precisely at the mobility edge. Our results demonstrate the richness of transport in quasiperiodic systems and reveal how specific spectral features (such as band edges, singular continuity, and mobility edges) govern their anomalous transport behavior.

The rest of this paper is organized as follows. In Sec.~\ref{secii}, we introduce the AAH model and detail its spectral properties, distinguishing between absolutely continuous, pure point, and singular continuous spectra. Methodologically, we present the transfer-matrix formalism and the nonequilibrium Green’s function framework for computing the energy-resolved Landauer conductance $\mathcal{G}(E,L)$, alongside the periodic approximation technique used to target specific eigenenergies with high precision. In Sec.~\ref{seciii}, we perform a comprehensive analysis of transport in the AAH model through finite-size scaling. In Subsec.~\ref{seciii}A (extended phase), we demonstrate the emergence of exceptional points at band edges, which leads to a universal $\mathcal{G}_{\text{typ}}\sim L^{-2}$ scaling. Subsec.~\ref{seciii}B discusses the exponential suppression of conductance in the localized phase. In Subsec.~\ref{seciii}C (critical phase), we reveal that the conductance exhibits log-periodic oscillations due to discrete scale invariance, and show how a ``stroboscopic'' analysis recovers a robust power-law scaling. Section~\ref{seciv} extends the analysis to the GAAH model which hosts a mixed spectrum. We specifically investigate transport across the mobility edge and provide numerical evidence that the edge itself acts as an insulating state located within a spectral gap. Finally, in Sec.~\ref{secv}, we summarize our main results and briefly comment on related open questions.

\section{Model and Method}\label{secii}
In this section, we first discuss the spectral properties of the AAH model across different phases in Subsection A. We then introduce the transfer-matrix approach and the associated Lyapunov exponent in Subsection B, which characterize the spatial localization properties of the eigenstates. Finally, Subsection C details the transport setup and the nonequilibrium Green’s function formalism employed to compute the conductance via the transfer matrix.
\subsection{AAH model and its spectrum}
The AAH model is a paradigmatic one-dimensional lattice model with a quasiperiodic onsite potential that is incommensurate with the lattice spacing. For a finite chain of length $L$, the Hamiltonian reads
\begin{align}{\label{formula:model}}
H = t \sum_{n=1}^{L-1} (c^\dagger_{n+1}c_n + \text{H.c.}) + \sum_{n=1}^{L} 2V\cos(2\pi bn+\theta)c^\dagger_n c_n.
\end{align}
Here $t$ is the nearest-neighbor hopping strength. $V$ controls the strength of the quasiperiodic modulation. In the following, we set the energy unit as $t=1$ and the potential strength $V\geq 0$. $b=(\sqrt{5}-1)/2$ is the inverse golden ratio (an irrational number). The quasiperiodic potential is deterministic and correlated, which evades the usual one-parameter scaling argument for uncorrelated random disorder. 

In the thermodynamic limit $L\rightarrow\infty$, the AAH model exhibits a well-known localization–delocalization transition at $V=1$. This is in contrast to the case of uncorrelated random disorder, where in one dimension, any finite disorder localizes single-particle states and hence no disorder-driven Anderson transition occurs~\cite{Abrahams1979}. The transition in the AAH model follows from its self-duality under a Fourier  transformation~\cite{Aubry1980,Gordon1997,Jitomirskaya1999}. For $V<1$, the system is in the extended phase and all eigenstates are extended; for $V>1$, the quasiperiodic modulation is sufficiently strong that all eigenstates become exponentially localized. Exactly at $V=1$, the system is critical and the eigenstates are commonly described as multifractal~\cite{Kohmoto1987,Hiramoto1989}. Because the incommensurate modulation frequency $b$ is irrational, the spectrum is a Cantor set \cite{Avila2009,Last1994}. Moreover, the spectral type depends crucially on $V$: in the extended phase, the spectrum is purely absolutely continuous, in the localized phase, it is pure point, and at the critical point, it is singular continuous~\cite{Jitomirskaya1999,Avila2015,Jitomirskaya1994}.

For these three distinct types of spectra, we need to further clarify their finite-size properties and precisely locate different spectral regions, which will become crucial in our later study of transport. In the extended phase, we distinguish three types of eigenenergies: those inside the spectrum ($E_{\text{in}}$), those outside the spectrum (i.e., in the gaps, $E_{\text{out}}$), and those at the band edges ($E_{\text{edge}}$). As we will demonstrate later, in the localized and critical phases the notion of a band edge is ill-defined, so we only distinguish between energies within the spectrum ($E_{\text{in}}$) and those outside it ($E_{\text{out}}$).

In practical calculations, we always work with finite systems and study how physical quantities approach the thermodynamic limit via finite-size scaling. For the AAH model, a common approach is the periodic-approximation technique~\cite{Kohmoto1987,Hiramoto1989}. The core idea is to replace the irrational number $b$ with a sequence of rational approximants $F_{n-1}/F_n$, where $F_n$ is the $n$th Fibonacci number. The approximation improves as $n$ increases, and the irrational value is recovered in the limit $b=\lim_{n\to\infty}F_{n-1}/F_n$. For any finite $n$, the model becomes periodic with period $F_n$. With periodic boundary conditions, Bloch's theorem applies and the spectrum consists of $F_n$ bands. A key feature of this hierarchical construction is that each band splits into three sub-bands at the next level of the approximation~\cite{Ostlund1983,Kohmoto1987,Hiramoto1989,Last1994}, which can be labeled as a central band ($0$) and two sidebands ($1$ and $\bar{1}$). This nesting allows any energy in the spectrum to be represented by an infinite symbolic sequence, or “code,” i.e., $\{s_1,s_2,\dots\}$, where each $s_i\in\{1,0,\bar{1}\}$. The structure of this code provides a detailed classification of energies. For instance, a code ending with an infinite sequence of zeros corresponds to an energy inside a band ($E_{\text{in}}$), while a code ending with an infinite sequence of ones (or $\bar{1}$s) corresponds to an energy at the global upper (or lower) edge of the spectrum ($E_{\text{edge}}$). Energies outside the spectrum, $E_{\text{out}}$, are given by the complement of these bands.

Since the coding itself is independent of the potential strength $V$, this symbolic classification may seem, at first glance, to conflict with our earlier, phase-dependent categorization of eigenenergies. The resolution lies in how the bandwidths $B_n$ (of the sub-bands in the $n$-th periodic approximation) behave in the thermodynamic limit ($n \to \infty$)~\cite{Hiramoto1989}. In the extended phase ($V<1$), the sub-band widths decay only algebraically with $n$. The spectrum is absolutely continuous with a finite total bandwidth (a ``fat" Cantor set), meaning the spectral clusters effectively form continuous bands with well-defined boundaries. In our practical calculations with finite $n$, we use exact diagonalization and select representative energies by choosing specific quasi-momenta $k$; for example, band-edge energies are typically found at the Brillouin zone boundaries ($k=0, \pi/F_n$), while band-interior energies can be found at other $k$’s. In contrast, in the localized phase ($V>1$), the sub-band widths shrink exponentially ($B_n \sim e^{-\eta F_n}$), causing the spectrum to collapse into a set of discrete points. The concept of a ``band" disappears, rendering the distinction between band-interior and band-edge states meaningless. Thus, we only distinguish between energies within ($E_{\text{in}}$) and outside ($E_{\text{out}}$) the spectrum. Similarly, at the critical point ($V=1$), although the sub-band widths decay as a power law $B_n \sim F_n^{-\gamma}$ (with $\gamma > 1$), the spectrum becomes singular continuous with zero Lebesgue measure. Strictly speaking, in the thermodynamic limit, the bands vanish into a fractal dust, and the distinction between band edges and band interiors ceases to exist.

\subsection{Transfer matrix and Lyapunov exponent}
Starting from the  eigenvalue equation $H|\psi\rangle=E|\psi\rangle$, the AAH model gives a three-term recursion relation for the site amplitudes $\psi_n$,
\begin{equation}
\psi_{n+1}+\psi_{n-1}+2V\cos(2\pi bn+\theta)\,\psi_n = E\,\psi_n .
\end{equation}
Introducing the two-component vector $\Phi_n=(\psi_n,\psi_{n-1})^{T}$, this recursion can be recast into a transfer-matrix form, $\Phi_{n+1}=T_n(E)\Phi_n$, where the single-site transfer matrix is
\begin{equation}
T_n(E)=
\begin{pmatrix}
E-2V\cos(2\pi bn+\theta) & -1\\
1 & 0
\end{pmatrix}.
\end{equation}
For a finite chain of length $L$, the full transfer matrix is the ordered product of the local matrices,
\begin{equation}\label{tmatrix}
T(E,L)=\prod_{n=1}^{L}T_n(E),
\end{equation}
which propagates the wavefunction amplitudes from one end of the chain to the other.

A central quantity extracted from $T(E,L)$ is the Lyapunov exponent $\lambda(E)$, which characterizes the asymptotic exponential growth rate of eigenstates. A standard definition is
\begin{equation}
\lambda(E)=\lim_{L\to\infty}\frac{1}{L}\ln \|T(E,L)\|,
\end{equation}
where $\|\cdot\|$ denotes the matrix norm. The Lyapunov exponent depends explicitly on $E$. When $E$ corresponds to a localized eigenstate, $\lambda(E)>0$ and the associated localization length is $\xi(E)=1/\lambda(E)$. By contrast, $\lambda(E)=0$ indicates the absence of exponential growth, consistent with extended or critical behavior. The energy dependence of $\lambda(E)$ provides a practical way to identify spectral regions in different phases of the AAH model. In the extended phase ($V<1$), $\lambda(E)$ vanishes for energies inside the spectrum, while it becomes positive in spectral gaps; the band edges $E_{\text{edge}}$ can be located by the sharp boundary between regions with $\lambda(E_{\text{in}})=0$ and $\lambda(E_{\text{out}})>0$. In the localized phase ($V>1$), $\lambda(E)$ is positive for all energies in the spectrum and directly sets the inverse localization length for eigenenergies $E_{\text{in}}$. At the critical point ($V=1$), $\lambda(E_{\text{in}})=0$ for all energies in the (singular-continuous) spectrum, whereas $\lambda(E_{\text{out}})>0$ for energies outside it. Note that the Lyapunov exponent is 
defined in the thermodynamic limit. In numerical calculations, we use the irrational value of $b$ without Fibonacci approximation. This involves iterating the transfer matrix multiplication until $\lambda(E)$ converges to a desired precision~\cite{Slevin2014}. One may periodically renormalize the transfer-matrix product (e.g., via QR or Gram--Schmidt re-orthogonalization) to prevent numerical overflow/underflow and ensure stable calculations for very large system sizes~\cite{MacKinnon1983,Markos2006,Slevin2014}.

\subsection{Landauer conductance}
We consider a finite quasiperiodic chain of length $L$ connected at its ends (sites $1$ and $L$) to two semi-infinite, noninteracting tight-binding leads. The leads serve as particle reservoirs at zero temperature, characterized by chemical potentials $\mu_L = E$ and $\mu_R = E-\delta E$. In the linear-response regime $\delta E\to 0$, the steady-state conductance is determined by the transmission at the reservoir chemical potential $E$ and can be quantified by the Landauer formula~\cite{Landauer1970,Economou1981}. Using the nonequilibrium Green's function, the zero-temperature conductance can be expressed as~\cite{DharSen2006,Purkayastha2016}:
\begin{align}\label{formula:G}
\mathcal{G}(E,L) = \frac{\mathcal{J}^2(E)}{2\pi|\det M(E,L)|^2}.
\end{align}
The conductance is expressed in units of $e^2/h$. Here, $\mathcal{J}(E)$ is the spectral function of the leads (assumed identical for the two ends), and $M(E,L) = E-H-\Sigma(E)$ is the inverse retarded Green's function, where $H$ is the $L\times L$ Hamiltonian of the isolated chain and $\Sigma(E)$ is the self-energy describing the influence of the leads. The self-energy is non-zero only at the boundaries: $\Sigma_{1,1}(E)=\Sigma_{L,L}(E) = -\frac{\gamma^2E}{2t_B^2}-i\frac{\gamma^2}{t_B}\sqrt{1-(\frac{E}{2t_B})^2}$, where $\gamma$ is the system-lead coupling strength and $t_B$ sets the bandwidth of the leads. The specific values of these microscopic parameters do not affect the universal scaling properties of the conductance discussed below.

Given the tridiagonal structure of the Hamiltonian $H$, the determinant $\det M(E,L)$ can be computed efficiently using the transfer matrix method~\cite{Saha2023}. Specifically, the determinant relates to the global transfer matrix $T(E,L)$ [defined in Eq. (\ref{tmatrix})] via the following relation:
\begin{align}\label{formula:G_TM}
    \det M(E,L) = \begin{pmatrix} 1 & -\Sigma_{1,1} \end{pmatrix} T(E,L) \begin{pmatrix} 1 \\ -\Sigma_{L,L} \end{pmatrix}.
\end{align}
We are primarily interested in the finite-size scaling behavior of the conductance rather than its specific value for a single finite realization, as scaling laws reveal universal transport properties in the thermodynamic limit ($L\to\infty$). Since the spectrum of the AAH model is independent of the phase offset $\theta$ in this limit due to ergodicity, we characterize the transport by the typical conductance, defined as the geometric mean over the phase $\theta$: $\mathcal{G}_{\text{typ}}(E,L) \equiv \exp(\langle\ln\mathcal{G}(E,L)\rangle_{\theta})$. This phase averaging is equivalent to sampling various segments of an infinite quasiperiodic chain. By utilizing the geometric mean (typical value) rather than the arithmetic mean, we effectively suppress rare resonant fluctuations, which is crucial for robustly extracting the underlying universal scaling exponents.

\section{Transport: the AAH model}\label{seciii}
In this section, we explore transport in the AAH model within three regimes corresponding to the extended ($V<1$), localized ($V>1$), and critical ($V=1$) phases. Our analysis centers on how the typical conductance $\mathcal{G}_{typ}(E,L)$ scales with system size $L$ as the bath chemical potential $E$ is varied.
\subsection{Extended phase}
\begin{figure}[!t]
    \centering
    \includegraphics[width=3.35in]{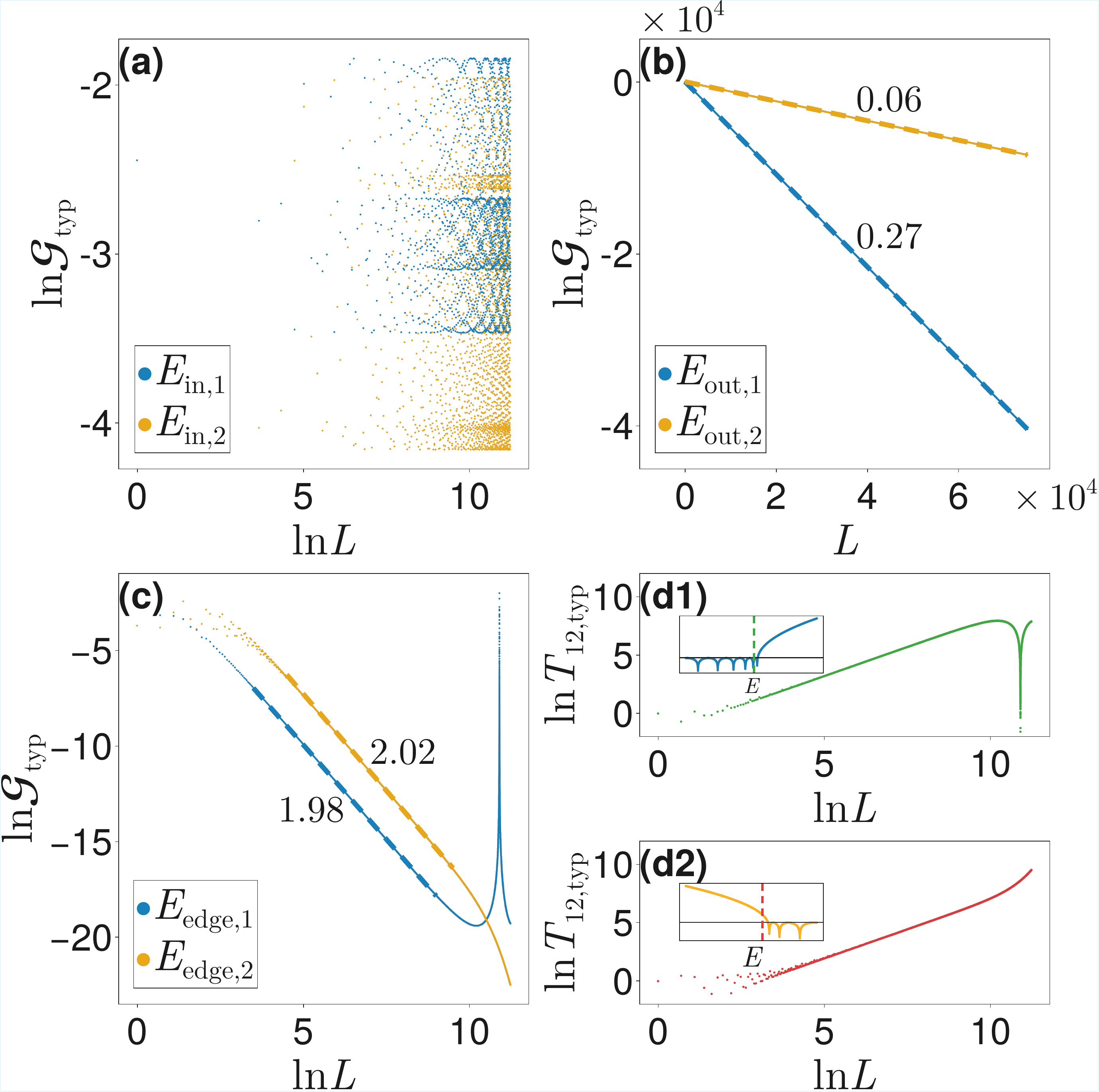}
\caption{\textbf{Transport in the extended phase of the AAH model.} (a)--(c) Typical conductance $\mathcal{G}_{\text{typ}}$ as a function of system size $L$ (up to Fibonacci number $F_{24}$) for the AAH model with $V=0.5$. The panels correspond to energies within the band ($E_{\text{in}}$), in the gap ($E_{\text{out}}$), and at the band edge ($E_{\text{edge}}$), respectively. Two representative energies are shown for each case, and data are geometrically averaged over 500 samples. Dashed lines indicate analytical fits using Eq.~\eqref{G_loc} in (b) and Eq.~\eqref{G_edge} in (c), where slopes are presented. Target energies at band edges were identified using the periodic approximation ($b \approx F_{20}/F_{21}$). The specific energy values are: $E_{\text{in}} \in \{0, 1.8398\}$, $E_{\text{out}} \in \{0.8163, 1.6816\}$, and $E_{\text{edge}} \in \{0.3351, 1.7433\}$. 
(d1)--(d2) Linear growth of the off-diagonal transfer matrix element $T_{12}$ of the transfer matrix [see its definition in Eq. (\ref{tmatrix})] at the two selected band-edge energies. The insets display the half-trace of the transfer matrix for $L=F_{24}$, with the positions of $E_{\text{edge}}$ marked by vertical dashed lines.}\label{fig1}
\end{figure}
We begin our analysis with the extended phase ($V<1$) of the AAH model and present the numerical results first. Using the transfer matrix defined in Eq.~(\ref{tmatrix}), we calculate the typical conductance $\mathcal{G}_{\text{typ}}(E,L)$ up to system sizes of $L=F_{24}$. Figures~\ref{fig1}(a-c) display the finite-size scaling of conductance when the bath chemical potential $E$ lies within a band ($E_{\text{in}}$), in a spectral gap ($E_{\text{out}}$), and at a band edge ($E_{\text{edge}}$), respectively. To locate these eigenenergies, we apply the periodic approximation discussed in Sec.~\ref{secii}A. We approximate the golden ratio using the rational approximant $b=F_{20}/F_{21}$ and select specific eigenenergies based on their symbolic sequence labeling. For band-interior energies, as shown in Fig.~\ref{fig1}(a), the typical conductance exhibits no systematic decay with system size ($\mathcal{G}_{\text{typ}}(E_{\textit{in}},L) \sim L^0$) but displays significant fluctuations. This behavior is characteristic of ballistic transport, where transmission is modulated by finite-size quantum interference effects. In contrast, for energies in the gap, we observe a clear exponential decay [note the semi-log scale in Fig.~\ref{fig1}(b)]. The data fits perfectly to:
\begin{align}\label{G_out}
\mathcal{G}_{\text{typ}}(E_{\text{out}},L)\sim e^{-2\lambda(E)L},
\end{align}
where $\lambda(E)$ is the Lyapunov exponent at energy $E$. This confirms the absence of transport at zero temperature when the chemical potential lies within a band gap.

Interestingly, when $E$ is selected at the band edges, the conductance exhibits a distinct power-law decay, up to a characteristic length set by the rational approximant of $b$. As we employ higher-order Fibonacci approximants, $b$ becomes closer to the golden ratio, and the selected $E_{\text{edge}}$ lies closer to the true band edge of the infinite quasiperiodic system. Consequently, the spatial range where this power-law scaling holds is enlarged. Further fitting of this decay to Eq.~\eqref{G_edge} yields:
\begin{align}\label{G_edge}
\mathcal{G}_{\text{typ}}(E_{\text{edge}},L)\sim L^{-2},
\end{align}
which corresponds to $\alpha=2$. In summary, as the bath chemical potential is tuned from the band interior across the edge, the transport undergoes a transition from ballistic to subdiffusive (exactly at the band edge), and finally to a regime of no transport.

To analyze the dependence of transport properties on energy $E$, we examine the conductance expressions in Eqs.~(\ref{formula:G}) and (\ref{formula:G_TM}), along with the structure of the transfer matrix in Eq.~(\ref{tmatrix}). In the limit of large system size ($L \gg 1$), universal transport behavior is expected to emerge. For a given energy $E$, the transfer matrix $T(E,L)$ has two eigenvalues, $\varepsilon_1$ and $\varepsilon_2$, satisfying $\varepsilon_1\varepsilon_2 = \det T(E,L) = 1$. Denoting $c(E) = \frac{1}{2}\mathrm{Tr}[T(E,L)]$, the eigenvalues can be written as $\varepsilon_{1,2} = c \pm \sqrt{c^2-1}$. Without loss of generality, we assume $|\varepsilon_1| \ge |\varepsilon_2|$. The Lyapunov exponent is then given by $\lambda = \frac{1}{L}\ln |\varepsilon_1|$. The value of $c(E)$ clearly distinguishes between different energy regimes. For energies within a band or at a band edge, the wavefunction is extended or multifractal with $\lambda=0$. It follows that the two eigenvalues are complex conjugates with unit modulus ($|\varepsilon_1|=|\varepsilon_2|=1$) and $|c(E)| \le 1$. Conversely, for energies in a spectral gap ($E_{\text{out}}$), the wavefunction is evanescent and $\lambda > 0$. This implies the eigenvalues are real and satisfy $|\varepsilon_1| > 1 > |\varepsilon_2|$, corresponding to the case $|c(E)| > 1$. As a result, the norm of the transfer matrix grows as $\|T(E,L)\| \sim e^{\lambda(E)L}$, leading to the exponential decay [see  Eq.~(\ref{G_out}) ] of the typical conductance observed in Fig. \ref{fig1}(b).

The band edge is special. It represents the boundary separating the above two regimes. When the energy $E$ lies at a band edge, $|c(E_{\text{edge}})| = 1$. Consequently, the two eigenvalues $\varepsilon_{1,2}$ of the transfer matrix become degenerate, with $\varepsilon_1=\varepsilon_2=\pm 1$. A matrix with degenerate eigenvalues falls into one of two categories. The first is the trivial case where the matrix is diagonalizable (e.g., proportional to the identity matrix), which would imply ballistic transport. The second is the nontrivial case where the matrix is non-diagonalizable and assumes a Jordan normal form. In this scenario, the eigenvalue possesses an algebraic multiplicity of two but a geometric multiplicity of one. This structure characterizes an exceptional point of the transfer matrix. Therefore, the $L^{-2}$ scaling of the conductance observed at the band edge corresponds to this second scenario, signaling the presence of an exceptional point.

It is important to note that the individual single-site transfer matrices $T_n(E)$ do not inherently possess a Jordan block structure; rather, the exceptional point is an emergent property of the full transfer matrix accumulated over large $L$. Such an emergent exceptional point is unique to the quasiperiodic nature of the system. While for any finite system the trace is analytic, in the thermodynamic limit, the band edge acts as a singular point where the trace touches the critical value $\pm 2$, effectively enforcing the Jordan block structure on the transfer matrix. This stands in sharp contrast to the case of simple periodic lattices \cite{Saha2023,Saha2025a}. In those systems, the transfer matrix is defined for a single unit cell, and the Jordan normal form appears naturally whenever the chemical potential aligns with a band edge. A direct consequence of this emergent EP is the linear growth of the transfer matrix elements with increasing system size $L$. This is illustrated in Figs.~\ref{fig1}(d1) and (d2), which track the evolution of the off-diagonal element $T_{12}$ of $T(E,L)$. The linear growth regime is clearly visible for system sizes that are sufficiently large yet remain below the characteristic length scale imposed by the rational approximant of $b$. This linear divergence of the matrix elements ultimately drives the $L^{-2}$ decay of the conductance, as dictated by Eq.~(\ref{formula:G}).

To understand the linear growth of the transfer matrix elements, we appeal to the properties of the system's rational approximants. Although we calculate the transfer matrix using the true irrational golden ratio, the energy $E_{\text{edge}}$ is located via a periodic approximation with some Fibonacci number $F_{n_0}$. For such a periodic system, the band edges are strictly defined by the condition $|\mathrm{Tr}[M_{Q}(E)]| = 2$. Excluding the trivial identity case, such a matrix is defective and is similar to a Jordan block:
\begin{equation}
M_Q(E_{\text{edge}}) = S J S^{-1} = S \begin{pmatrix} \eta & 1 \\ 0 & \eta \end{pmatrix} S^{-1},
\end{equation}
where $\eta = \pm 1$. Physically, this implies that at the band edge, the wavefunction amplitude exhibits algebraic growth rather than simple oscillation (extended) or exponential decay (localized). For a system of size $L$ comprising roughly $n$ such periods ($L \approx nQ$), the global transfer matrix behaves effectively as the $n$-th power of this block. Utilizing the property $J^n = \begin{pmatrix} \eta^n & n\eta^{n-1} \\ 0 & \eta^n \end{pmatrix}$, we obtain:
\begin{equation}
T(E_{\text{edge}}, L) \sim [M_Q]^n \sim \begin{pmatrix} \dots & \sim n \\ 0 & \dots \end{pmatrix}.
\end{equation}
This derivation explains why the off-diagonal elements grow linearly with the number of effective periods $n$, and consequently with the system size $L$.

We next address the impact of finite precision in our numerical energy selection. Since $E_{\text{edge}}$ is determined numerically, the chosen energy may fundamentally correspond to an $E_{\text{in}}$ (band) or $E_{\text{out}}$ (gap) state that lies extremely close to the true critical edge. This proximity leads to finite-size crossover effects governed by diverging length scales~\cite{Hiramoto1992}:

\begin{itemize}
    \item If the chosen energy lies slightly inside the band ($E_{\text{in}}$), the group velocity $v_g$ vanishes as one approaches the edge ($v_g \to 0$). Physically, a vanishing group velocity implies that an infinitely long time (or length) is required to resolve the ballistic nature of the transport. Consequently, the system exhibits critical subdiffusive scaling ($\sim L^{-2}$) up to a large characteristic length $L_c \propto 1/v_g$, beyond which normal ballistic transport ($\sim L^0$) is recovered~\cite{Sutherland1987}.
    
    \item Conversely, if the energy lies slightly outside the band ($E_{\text{out}}$), the localization length $\xi$ diverges at the edge ($\xi \to \infty$). The system essentially "looks" critical up to a scale $L_c \sim \xi$. Only when the system size exceeds this correlation length ($L > L_c$) does the exponential decay become apparent.
\end{itemize}

This physical picture explains the robustness of the power-law scaling observed in Fig.~\ref{fig1}(c) as well as the departure beyond some length scale. Finally, regarding the thermodynamic limit, one might postulate that the incommensurate modulation acts as a perturbation $\Delta$, such that $T(L) \approx J(L) + \Delta$. However, the exact band edge represents a critical fixed point of the renormalization group flow associated with the trace map dynamical system~\cite{Kohmoto1983, Ostlund1983}. This critical manifold is marginally stable; the strict condition of being \textit{at} the phase boundary forbids the system from flowing into either the localized or extended regimes. Thus, the linear algebraic scaling characteristic of the Jordan block is preserved as the asymptotic universal behavior of the emergent exceptional point.

\subsection{Localized phase}
\begin{figure}[!t]
    \centering
    \includegraphics[width=3.35in]{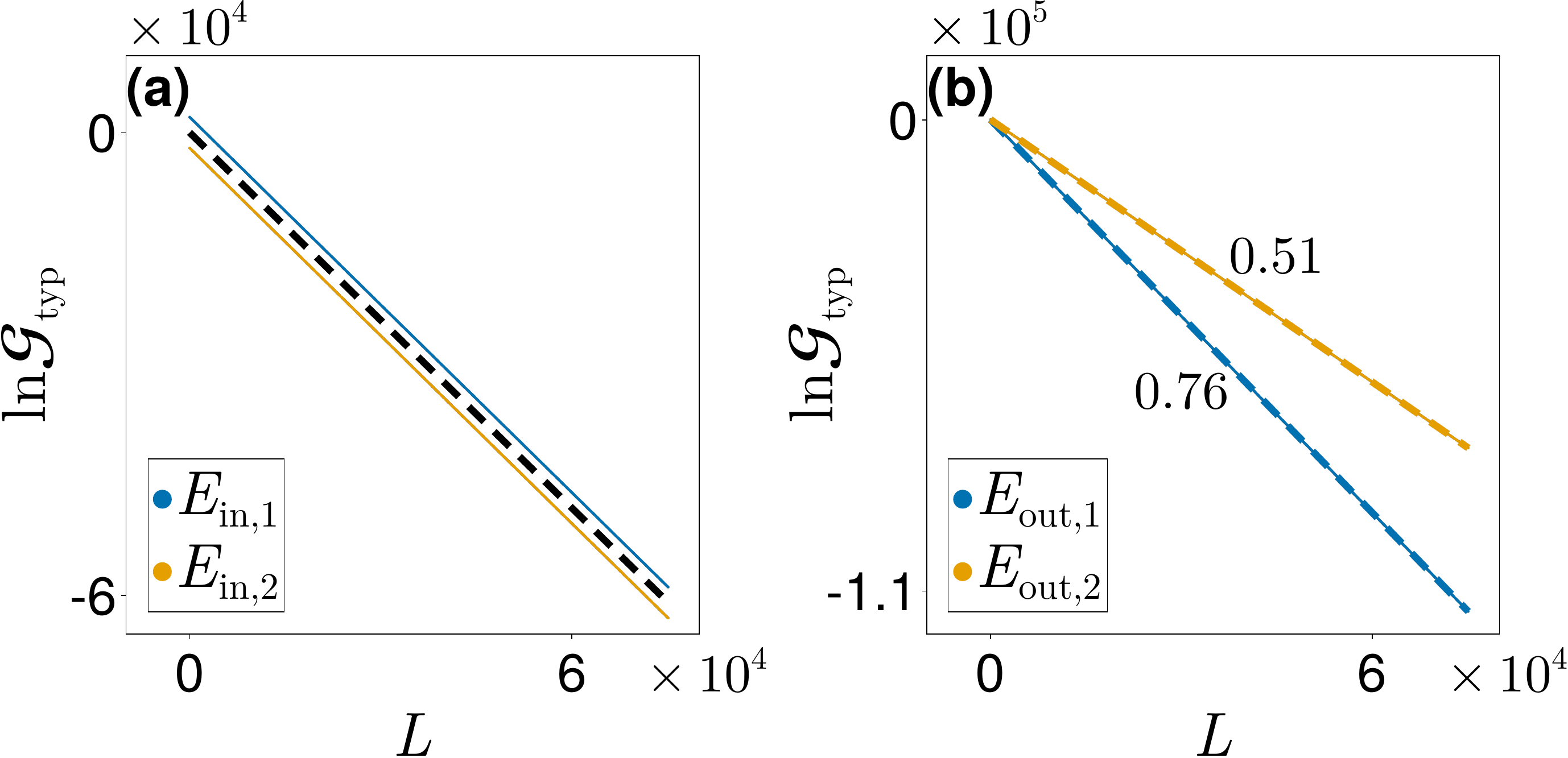}
\caption{\textbf{Transport in the localized phase of the AAH model.} Finite-size scaling of the typical conductance $\mathcal{G}_{\text{typ}}$ as a function of system size $L$ (up to $F_{24}$) is performed. $V=1.5$. (a) Energies within the spectrum ($E_{\text{in}}$). The data for different energies are slightly shifted vertically for clarity. The decay rate is governed by the analytic Lyapunov exponent $\lambda = \ln V$. (b) Energies in the gaps ($E_{\text{out}}$). The decay rate is governed by the energy-dependent Lyapunov exponent calculated numerically. Two representative energies are presented for each case, and data are geometrically averaged over 500 samples. Dashed lines denote exponential fits according to Eq.~\eqref{G_loc}. For $E_{\text{out}}$, the Lyapunov exponents are computed via the transfer matrix method with 5000 samples and $\sigma=0.0003$. The specific energy values used in the plot are: $E_{\text{in}} \in \{-0.0001, 0.3834\}$ and $E_{\text{out}} \in \{1.3179, 2.7339\}$.}\label{fig2}
\end{figure}
In the localized phase ($V>1$), the system hosts a pure point spectrum, corresponding to a complete set of exponentially localized eigenstates. A defining characteristic of this phase is that the Lyapunov exponent $\lambda(E)$ remains strictly positive for all energies, whether they lie within the allowed bands ($E_{\text{in}}$) or in the spectral gaps ($E_{\text{out}}$). This stands in sharp contrast to the extended phase, where the Lyapunov exponent vanishes on the spectrum. For energies within the spectrum, the inverse Lyapunov exponent gives the localization length, $\xi = 1/\lambda(E)$. The positivity of the Lyapunov exponent implies that the transfer matrix elements grow exponentially with system size $L$. Consequently, the typical conductance decays exponentially for all energies:
\begin{align}\label{G_loc}
\mathcal{G}_{\text{typ}}(E,L) \sim e^{-2\lambda(E)L}.
\end{align}

For the AAH model, the Lyapunov exponent  $\lambda(E)$ within the spectrum can be obtained via Avila's global theory of one-frequency Schrödinger operators~\cite{Avila2015}. This theory establishes that for energies in the spectrum,  $\lambda(E)$ is determined solely by the strength of the quasiperiodic modulation, $\lambda(E) = \ln V$. Notably, the decay rate is independent of the specific energy $E$ as long as it lies within the bands. Our numerical results for the typical conductance are presented in Fig.~\ref{fig2}. The energies used in these calculations were identified using the periodic approximation method. As shown in Fig.~\ref{fig2}(a), when the chemical potential is tuned within the bands, the conductance curves for different energies are parallel, confirming that the exponential decay rate is independent of the bath chemical potential (universally $\approx 2\ln V$). In contrast, when $E$ lies inside a spectral gap, the decay rate is governed by the $E$-dependent Lyapunov exponent and thus varies with $E$, as illustrated in Fig.~\ref{fig2}(b). Therefore, in the localized phase, the system behaves as a perfect insulator with suppressed transport in the thermodynamic limit.

\subsection{Critical phase}
\begin{figure}[!t]
    \centering
    \includegraphics[width=3.35in]{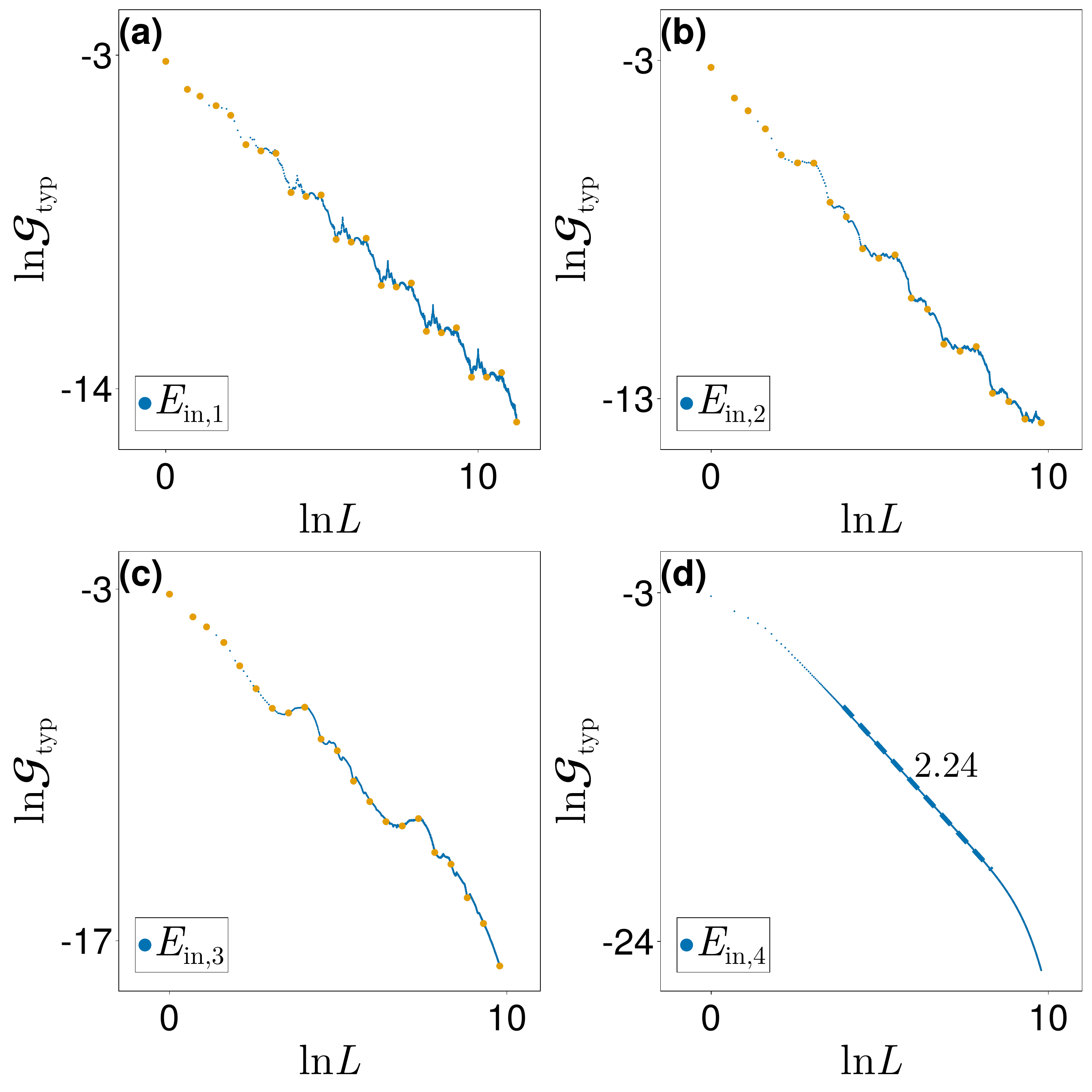}
\caption{\textbf{Transport in the critical phase of the AAH model.} Finite-size scaling of the typical conductance $\mathcal{G}_{\text{typ}}$ at the critical point ($V=1$) is shown. Four representative in-band energies ($E_{\text{in}}$) are analyzed. Data are geometrically averaged over 5000 samples. The maximum system size is $L=F_{24}$ for $E_{\text{in},1}$ and $L=F_{21}$ for the other cases. Data points corresponding to Fibonacci lengths ($L=F_n$) are highlighted with yellow dots. For $E_{\text{in},4}$, a power-law fit yields an exponent $\alpha \approx 2.24$ (dashed line). The specific energies and their corresponding symbolic codes are: (a) $E_{\text{in},1}=0$ ($\{000000\dots\}$);  (b) $E_{\text{in},2}=0.1719$ ($\{010101\dots\}$); (c) $E_{\text{in},3}=0.1874$ ($\{011011\dots\}$); and (d) $E_{\text{in},4}=0.1891$ ($\{011111\dots\}$).}
\label{fig3}
\end{figure}
In the AAH model, the critical point separating the delocalized and localized phases occurs at $V=1$. At this transition point, the energy spectrum forms a Cantor set of zero Lebesgue measure, and the eigenstates exhibit multifractal characteristics. Our numerical results for the typical conductance, calculated at four representative bath chemical potential $E$, are presented in Fig.~\ref{fig3}. To select these energies, we utilize the periodic approximation method and assign a distinct symbolic code to each. At first glance, the conductance scaling appears non-universal and lacks a smooth functional form; the data is heavily modulated with respect to system size. While some cases (e.g., $E_{\text{in},4}$) suggest an underlying power-law decay. A closer inspection reveals a general feature: the conductance curves are superimposed with regular peaks that align with system sizes corresponding to Fibonacci numbers ($L=F_n$). %The presence of these oscillations indicates that the system possesses discrete, rather than continuous, scale invariance.

This behavior is identified as log-periodic oscillation, a hallmark of systems possessing discrete scale symmetry~\cite{Sornette1998}. These oscillations arise in systems where the continuous scale invariance is broken into a discrete subgroup, leading to complex critical exponents~\cite{Sornette1998, Gluzman2002}. The periodicity of these oscillations on a logarithmic scale is governed by an energy-dependent inflation factor. To understand this, we recall the recursive splitting of energy bands under the real-space renormalization group scheme~\cite{Kohmoto1983, Ostlund1983}. The symbolic code essentially dictates the trajectory of an energy state as one ``zooms out'' the spectrum (renormalization), which is equivalent to rescaling the system size.

\begin{figure}[!t]
    \centering
    \includegraphics[width=3.35in]{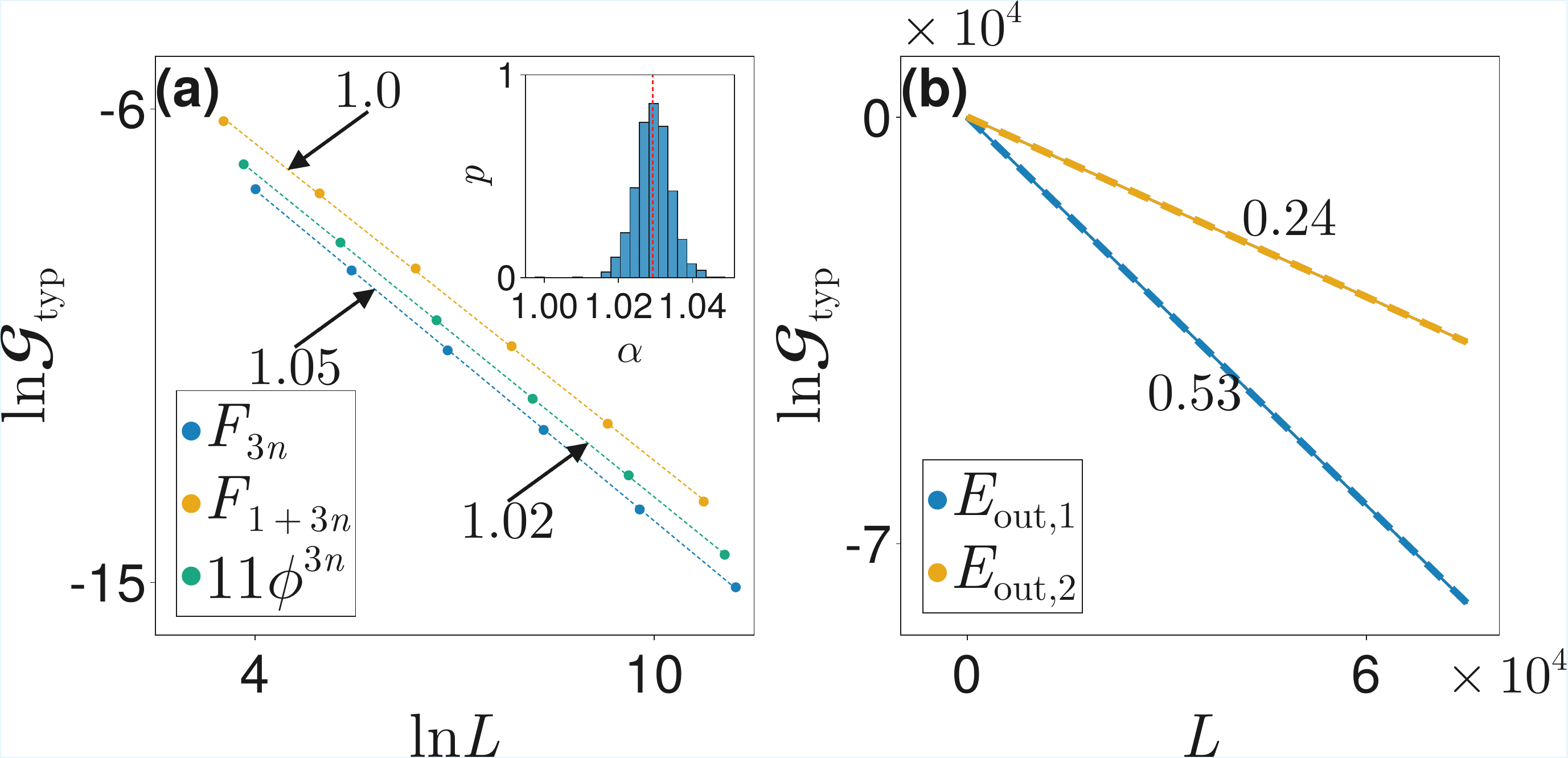}
  \caption{\textbf{Stroboscopic analysis and gap transport in the critical phase.} (a) Stroboscopic finite-size scaling of the typical conductance for the band-center state $E_{\text{in},1}$. The conductance is plotted for selected subsequences of system sizes that share the same inflation factor ($\lambda \approx \phi^3$): $L \in \{F_{3n}\}$, $\{F_{3n+1}\}$, and a geometric progression $L \approx 11\phi^{3n}$. Dashed lines indicate power-law fits according to Eq.~\eqref{G_crit}. \textit{Inset:} Probability distribution of the scaling exponent $\alpha$, obtained from an ensemble of length subsequences satisfying the $\phi^3$ scaling relation. Each fit utilizes at least three data points. The distribution centers around a mean value $\langle \alpha \rangle \approx 1.03$. (b) Exponential decay of the typical conductance for energies within spectral gaps ($E_{\text{out}}$). Dashed lines represent fits to Eq.~\eqref{G_loc}, where the Lyapunov exponents are calculated numerically with $\sigma=0.0003$ and 5000 samples. The specific gap energies are $E_{\text{out},1}=1.0316$ and $E_{\text{out},2}=2.1580$.}
\label{fig4}
\end{figure}

Crucially, the scaling factor in the thermodynamic limit is uniquely determined by the asymptotic sequence of the symbolic code. For instance, the central band (code $0$) is known to be self-similar after three renormalization steps~\cite{Kohmoto1983}, contributing an inflation factor of $ \lim_{n\to\infty} F_{n+3}/F_n = \phi^3$, where $\phi$ is the golden ratio. A side band (code $1$ or $\bar{1}$) is self-similar after two steps, contributing a factor of $\lim_{n\to\infty} F_{n+2}/F_n = \phi^2$ ($\phi=1/b$ is the golden ratio). This renormalization picture successfully explains the observed log-periodic oscillations in Figs.~\ref{fig4}(a)-(c) with the correct inflation factors: $\phi^3$ for $E_{\text{in},1}$ with code $\{\dots000000\dots\}$, $\phi^5$ for $E_{\text{in},2}$ with code $\{\dots010101\dots\}$, and $\phi^7$ for $E_{\text{in},3}$ with code $\{\dots011011\dots\}$. However, this simplified picture does not hold for $E_{\text{in},4}$ with code $\{\dots 111111 \dots\}$, where Fig.~\ref{fig4}(d) instead shows a continuous power-law scaling with an exponent $\alpha\approx 2.24$. We speculate that this anomaly stems from the distinction between ``bulk'' and ``edge'' fractal states, analogous to observations in the extended phase.

Motivated by the underlying discrete scale symmetry, we employ a ``stroboscopic'' analysis of the conductance. For a target energy $E_{\text{in}}$, we evaluate the conductance exclusively at a sequence of system sizes $L_k$ determined by its specific inflation factor (e.g., $L_{k+1} \approx \phi^3 L_k$). As presented in Fig.~\ref{fig4}(a), this procedure unveils a robust power-law scaling:
\begin{align}\label{G_crit}
\mathcal{G}_{\text{typ}}(E_{\text{in}},L) \sim L^{-\alpha(E_{\text{in}})}.
\end{align}
The scaling exponent $\alpha(E_{\text{in}})$ associated with a specific in-band energy depends on the chosen subsequence of lengths. Distinct subsequences yield slightly varying exponents, as evident in Fig.~\ref{fig4}(a). A statistical analysis performed over an ensemble of such sequences yields a Gaussian distribution with a mean value $\langle\alpha\rangle \approx 1.03$, indicating that the transport is (when the bath chemical potential is set here), on average, slightly subdiffusive. In principle, this analysis can be applied to every individual eigenenergy within the spectrum, highlighting the crucial impact of discrete scale symmetry on transport properties. We note the difference between our energy-resolved characterization and previous studies on transport in the critical phase. Prior investigations have typically characterized transport via wave packet dynamics~\cite{Hiramoto1988,Ketzmerick1997} or high-temperature steady currents~\cite{Varma2017}. In those setups, the observed transport exponents (e.g., the spreading exponent of a wave packet) arise from the weighted contributions of all eigenstates. Consequently, they correspond to an averaged effect over the singular continuous spectrum. In contrast, our approach resolves the transport properties of specific fractal eigenstates, revealing that the ``averaged'' diffusive-like behavior at the critical point from a statistical distribution of distinct scaling exponents calculated by varies sizes of samples.

Finally, when the bath chemical potential $E$ lies outside the spectrum, the transport is suppressed. The Lyapunov exponent is strictly positive, implying that the transfer matrix elements grow exponentially with system size. This is confirmed by our numerical data in Fig.~\ref{fig4}(b), which exhibits a clear exponential decay of the typical conductance. The scaling is thus described by Eq.~\eqref{G_out}, signifying a complete absence of transport in the gaps.

\section{Transport: the GAAH model}\label{seciv}
In this section, we extend our investigation to the GAAH model. This model was originally proposed by Ganeshan \textit{et al.}~\cite{Ganeshan2015}, with the Hamiltonian defined as:
\begin{align}\label{GAAH}
H = t\sum_{n=1}^{L-1}(c^\dagger_{n+1}c_n+\text{H.c.}) + \sum_{n=1}^{L}\frac{2V\cos(2\pi bn+\theta)}{1-U\cos(2\pi bn+\theta)}c^\dagger_nc_n.
\end{align}
Here, $V$ represents the strength of the quasiperiodic potential, and $b$ is set to the golden ratio. The additional parameter $U \in (-1, 1)$ introduces a mobility edge into the system. Unlike the standard AAH model, which exhibits a ``pure'' spectrum (either purely absolutely continuous or purely pure point depending on $V$), the GAAH model hosts a \textit{mixed} spectrum due to the analytic form of its potential. In the standard AAH model, self-duality occurs strictly at the critical point $V=1$. However, in the GAAH model, the concept of self-duality is generalized; the duality transformation maps the extended fraction of the spectrum onto the localized fraction, allowing for the coexistence of both phases separated by a precise energy boundary.

\begin{figure}[!t]
    \centering
    \includegraphics[width=3.35in]{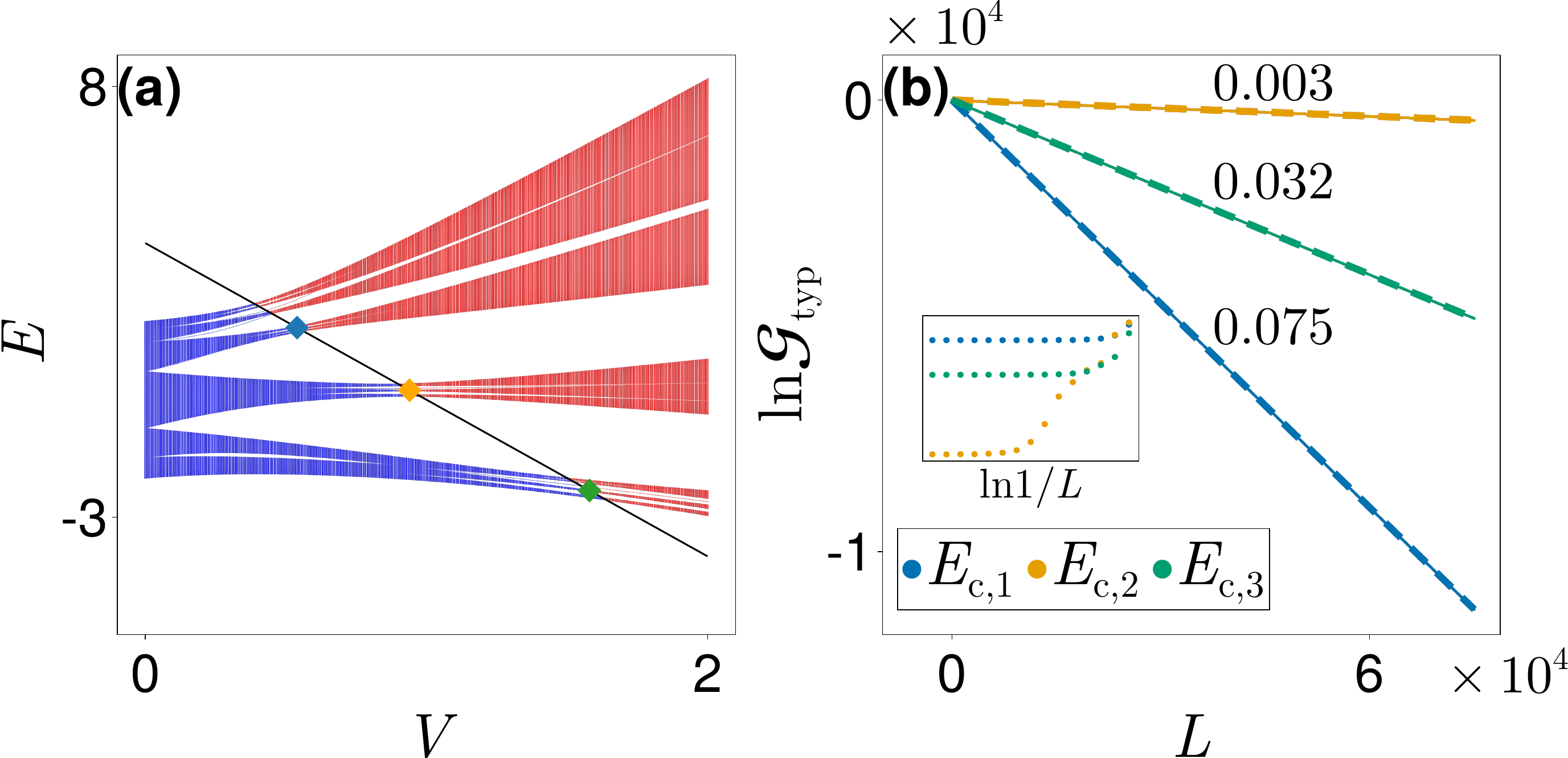}
\caption{\textbf{Spectrum and mobility edge transport in the GAAH model.} (a) Energy spectrum as a function of the quasiperiodic potential strength $V$, calculated via exact diagonalization. Parameters are set to $U=0.5$, $\theta=0$, and $b=(\sqrt{5}-1)/2$. Blue regions correspond to the absolutely continuous spectrum hosting extended states, while red regions correspond to the pure point spectrum hosting localized states. The solid black line marks the analytical mobility edge. (b) Finite-size scaling of the typical conductance $\mathcal{G}_{\text{typ}}$ for three distinct energies located exactly at the mobility edge. Results are geometrically averaged over 500 samples. Dashed lines indicate exponential fits according to Eq.~\eqref{G_out}, using Lyapunov exponents computed numerically ($\sigma=0.0003$, 5000 samples). \textit{Inset:} Finite-size scaling of the typical spectral gap $\Delta_{\text{typ}}$ in the vicinity of the selected mobility edge energies. The specific energy values are $E_{c,1}=1.84$, $E_{c,2}=0.24$, and $E_{c,3}=-2.32$.}
\label{fig5}
\end{figure}
\begin{figure}[!t]
    \centering
    \includegraphics[width=3.35in]{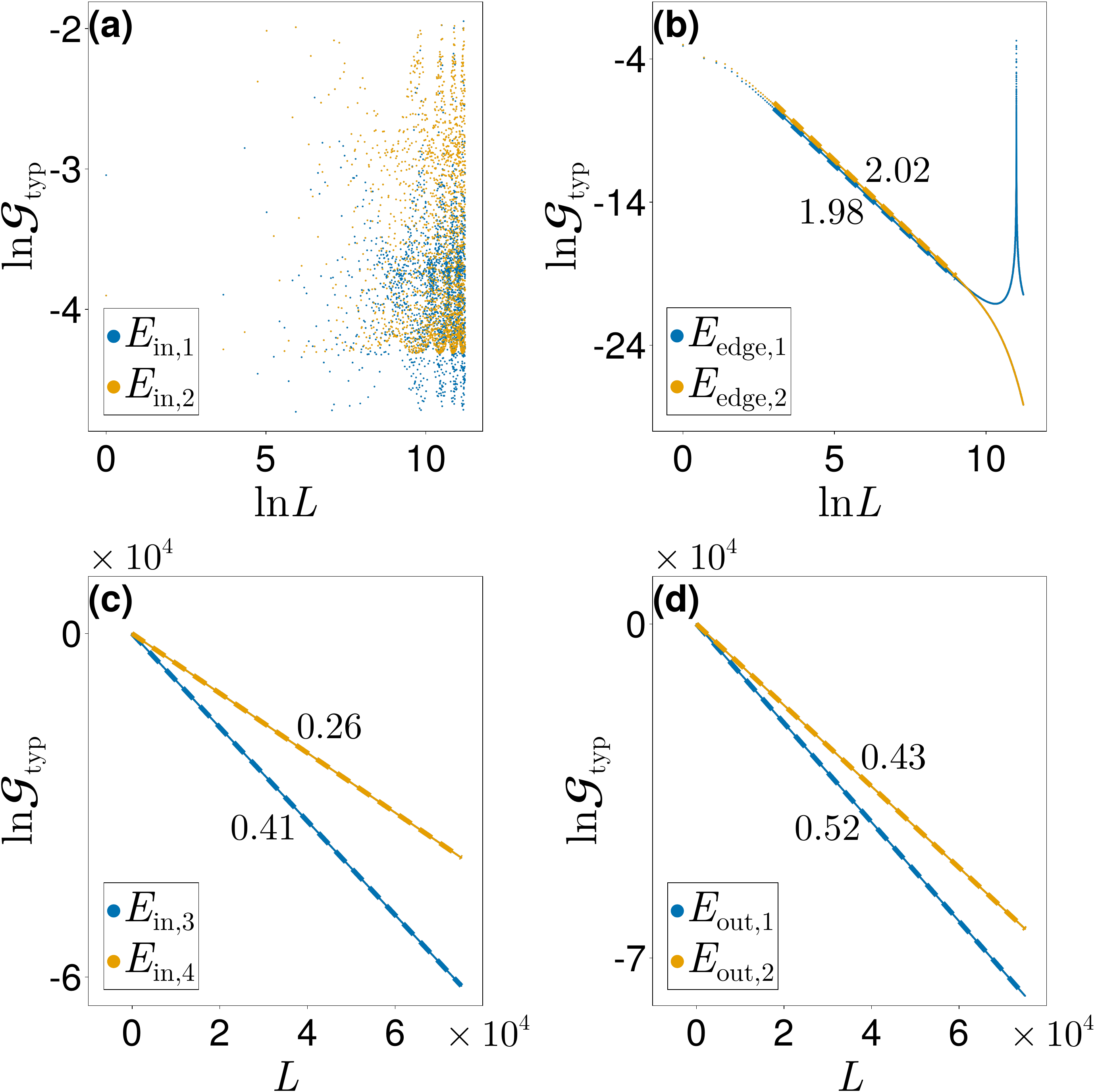}
\caption{\textbf{Finite-size scaling of the typical conductance $\mathcal{G}_{\text{typ}}$ for the GAAH model.} Parameters are $V=0.75$ and $U=0.5$, corresponding to a mobility edge at $E_c=1$. The maximum system size is $L=F_{24}$. (a) Band-interior energies in the extended regime ($E < E_c$), showing ballistic-like fluctuations.
(b) Band-edge energies in the extended regime ($E < E_c$), showing $L^{-2}$ (subdiffusive) scaling. Dashed lines indicate fits to Eq.~\eqref{G_edge}. (c) Band-interior energies in the localized regime ($E > E_c$), showing exponential decay. (d) Energies within spectral gaps ($E_{\text{out}}$), showing exponential decay. Results are geometrically averaged over 500 samples. Target energies were identified via periodic approximation ($b \approx F_{20}/F_{21}$). Dashed lines in (c) and (d) represent exponential fits [Eq.~\eqref{G_loc}] using numerically calculated Lyapunov exponents ($\sigma=0.0003$, 5000 samples). Specific energy values are: (a) Extended bulk $E_{\text{in}} \in \{0.2381, -1.6935\}$; (b) Extended edge $E_{\text{edge}} \in \{0.4150, 0.0602\}$; (c) Localized bulk $E_{\text{in}} \in \{2.8006, 2.0503\}$; and (d) Gaps $E_{\text{out}} \in \{1.1758, -0.6301\}$.}
\label{fig6}
\end{figure}
The mobility edge (ME) for this model can be rigorously derived from the self-duality condition~\cite{Ganeshan2015}. It partitions the energy spectrum into regions of absolutely continuous (extended) and pure point (localized) spectra. The exact expression for the mobility edge is given by:
\begin{align}
E_{\text{c}} = \frac{2(1-|V|)}{U}.
\end{align}
In Fig.~\ref{fig5}(a), we plot the energy spectrum as a function of $V$. The mobility edge $E_c$ appears as a distinct line separating two regimes. For energies below the ME line ($E < E_c$, indicated in blue), the spectrum is absolutely continuous, hosting extended states. Conversely, for energies above the ME line ($E > E_c$, indicated in red), the spectrum is pure point, consisting of localized states.
\begin{table*}[t]
\caption{\label{tab:summary} Summary of the typical conductance $\mathcal{G}_{\text{typ}}(E, L)$ for the AAH and GAAH models. The transport is determined by the specific region of the energy spectrum where the bath chemical potential $E$ is located. $\lambda$ is the Lyapunov exponent. $E_c$ is the level of mobility edge.}
\begin{ruledtabular}
\begin{tabular}{ccccc}
\textbf{Model} & \textbf{Phase} & \textbf{Spectral Region} & \textbf{Physical Picture} & \textbf{Scaling Law} \\ \hline
\multirow{3}{*}{AAH} & \multirow{3}{*}{\shortstack{Extended\\($V<1$)}} & Band Interior ($E_{\text{in}}$) & Ballistic Transport & $L^0$ (Fluctuating) \\
 & & Band Edge ($E_{\text{edge}}$) & Emergent EP / Subdiffusive & $ L^{-2}$ \\
 & & Band Gap ($E_{\text{out}}$) & Insulating / Evanescent & $ e^{-2\lambda L}$ \\ \hline
\multirow{2}{*}{AAH} & \multirow{2}{*}{\shortstack{Critical\\($V=1$)}} & Band Interior ($E_{\text{in}}$) & Multifractal / Singular continuous & \text{log-periodic, stroboscopically}~$L^{-\alpha(E)}$ \\
 & & Band Gap ($E_{\text{out}}$) & Insulating & $ e^{-2\lambda L}$ \\ \hline
AAH & Localized ($V>1$) & Entire Spectrum & Anderson Localization & $e^{-2\lambda L}$ \\ \hline
\multirow{3}{*}{GAAH} & \multirow{3}{*}{\shortstack{Mixed\\Phase}} & Extended Region ($E < E_c$) & Same as AAH Extended & $L^0$ or $L^{-2}$ \\
 & & Localized Region ($E > E_c$) & Same as AAH Localized & $e^{-2\lambda L}$ \\
 & & Mobility Edge ($E = E_c$) & Hidden Gap / Insulating & $e^{-2\lambda L}$ \\
\end{tabular}
\end{ruledtabular}
\end{table*}

Based on this spectral structure, we can adapt our previous categorization to the GAAH model. Below the ME line, the extended regime contains band-interior energies ($E_{\text{in}}$), band edges ($E_{\text{edge}}$), and spectral gaps ($E_{\text{out}}$). Above the ME line, the localized regime similarly comprises eigenenergies ($E_{\text{in}}$) and gaps ($E_{\text{out}}$). Since the typical conductance $\mathcal{G}_{\text{typ}}(E,L)$ is fundamentally determined by the local spectral nature of the selected energy $E$, we expect the transport laws identified in the AAH model to remain valid in the corresponding regions of the GAAH model. This expectation is fully verified by our numerical results shown in Fig.~\ref{fig6}. We again employ the transfer matrix method, selecting energies via the periodic approximation.
\begin{itemize}
    \item In Fig.~\ref{fig6}(a), the chemical potential is tuned to a band-interior energy below the ME line ($E < E_c$). The conductance shows no systematic decay ($\mathcal{G}_{\text{typ}} \sim L^0$) but exhibits large fluctuations, characteristic of ballistic transport.
    \item In Fig.~\ref{fig6}(b), selecting $E$ at band edges below the ME line reproduces the universal subdiffusive scaling $\mathcal{G}_{\text{typ}} \sim L^{-2}$. This confirms that the band edges in the extended phase of the GAAH model also host emergent exceptional points (of the transfer matrix) in the thermodynamic limit.
    \item In Fig.~\ref{fig6}(c), the energy is chosen within the spectrum but above the ME line ($E > E_c$). As expected for localized eigenstates, the conductance follows an exponential decay $\mathcal{G}_{\text{typ}} \sim e^{-2\lambda(E)L}$, where $\lambda(E)$ is the Lyapunov exponent.
    \item Finally, in Fig.~\ref{fig6}(d), we select energies inside the spectral gaps (both below and above the ME line). Since there are no available states at these energies, the wavefunctions are evanescent. Consequently, we observe a clear exponential decay of the conductance, signifying the absence of transport.
\end{itemize}

A crucial question remains regarding the nature of transport precisely \textit{at} the mobility edge, $E = E_{\text{c}}$. Intuitively, one might expect $E_{\text{c}}$ to represent a critical conducting state, as it marks the transition boundary in the spectrum (see Fig.~\ref{fig5}(a)). However, our numerical simulations in Fig.~\ref{fig5}(b) reveal a striking result: the conductance at $E_{\text{c}}$ consistently decays exponentially for all tested parameters. This insulating behavior suggests that the energy $E_c$ effectively behaves as if it lies within a spectral gap ($E_{\text{out}}$). To clarify this, we analyzed the typical spectral gap $\Delta_{\text{typ}}$ in the immediate vicinity of $E_{\text{c}}$ using exact diagonalization. The inset of Fig.~\ref{fig5}(b) presents the finite-size scaling of this gap. For all parameters, $\Delta_{\text{typ}}$ extrapolates to a finite, albeit small, value in the thermodynamic limit ($L\to\infty$). Based on this strong numerical evidence, we conjecture that the analytical mobility edge $E_{\text{c}}$ in the GAAH model generically falls within a spectral gap. Despite being the separation between phases, $E_{\text{c}}$ itself is not a conducting state but resides in a forbidden energy region. The mobility edge in quasiperiodic models should be understood strictly as a phase boundary in parameter space, which does not necessarily host a critical eigenstate itself.

\section{Conclusion and discussion}\label{secv}
To summarize, we have presented a systematic study of transport phenomena in the AAH model and the GAAH model with mobility edges. By combining high-precision methods for locating specific eigenenergies (via periodic approximation) with transfer matrix calculations of the Landauer conductance, we were able to reveal universal finite-size scaling laws that govern transport at large length scales. We demonstrate that the transport behavior is strictly dictated by the local spectral character at the level of the bath chemical potential. These universal scaling laws across different phases and spectral regions are summarized in Table~\ref{tab:summary}.

Our work yields three primary findings regarding the interplay between spectral geometry and quantum transport:

First, when the bath chemical potential is tuned to the band edge of an absolutely continuous spectrum, we observe the emergence of exceptional points in the global transfer matrix. This leads to a universal subdiffusive transport characterized by a $\mathcal{G}_{\text{typ}} \sim L^{-2}$ scaling. This phenomenon is a direct consequence of the continuity of the transfer matrix's trace: as the energy transitions from the band interior to the gap, the trace must constrain itself to $|\text{Tr}|=2$ at the boundary, enforcing a Jordan block structure.

Second, at the critical point of the AAH model, we demonstrated that the conductance exhibits log-periodic oscillations arising from the underlying discrete scale invariance. By performing a ``stroboscopic'' analysis consistent with the energy-dependent inflation factors, we identified a robust power-law scaling. 

Third, for the GAAH model, we reported that transport is suppressed exponentially exactly at the mobility edge. Our numerical evidence suggests that the analytical mobility edge is located within a spectral gap, rendering it an insulating point rather than a critical conducting state. While we have confirmed this for several parameter sets, proving this conjecture rigorously and even for all self-dual models of this type remains an important open problem for mathematical physics.

These results underscore the intrinsic link between quantum transport and local spectral properties in quasiperiodic systems. We expect that our results can be generalized to other quasiperiodic systems, including those with long-range hopping~\cite{Saha2025a,Gopalakrishnan2017} or other irrational modulation ratios. Furthermore, our findings in the Hermitian domain may offer insights for non-Hermitian and open quasiperiodic systems. Recent developments in non-Hermitian and open quantum systems reveal that the interplay between spectrum and transport is even more profound when unitarity is broken~\cite{WangPRL,Xing2025,Hu_PRB,JD_NP}. For instance, in quasiperiodic non-Hermitian systems, the complex nature of the spectrum has been shown to dictate novel universal scaling laws in spreading dynamics~\cite{Xing2025}. Specifically, the spreading exponent is directly related to the imaginary density of states at the band tail. Since standard non-equilibrium Green's function methods are typically inapplicable in non-unitary regimes, it would be interesting to investigate how the interplay between quasiperiodicity and non-Hermiticity affects the spectral geometry and the resulting transport properties.

\section{acknowledgments}
This work is supported by the National Key Research and Development Program of China (Grants No. 2022YFA1405800 and No. 2023YFA1406704) and National Natural Science Foundation of China (Grant No. 12474496 and No. 12547107).

\end{document}